\documentclass[amsfonts,prd,aps]{revtex4}

\usepackage{graphicx}

\begin{document}

\title{Quantum scalar field in quantum gravity:\\ the vacuum in the
spherically symmetric case}

\author{ Rodolfo Gambini$^{1}$,
Jorge Pullin$^{2}$,
Saeed Rastgoo$^{1}$} 
\affiliation {
1. Instituto de F\'{\i}sica, Facultad de Ciencias, 
Igu\'a 4225, esq. Mataojo, Montevideo, Uruguay. \\
2. Department of Physics and Astronomy, Louisiana State University,
Baton Rouge, LA 70803-4001}

\begin{abstract}
  We study gravity coupled to a scalar field in spherical symmetry
  using loop quantum gravity techniques. Since this model has local
  degrees of freedom, one has to face ``the problem of dynamics'',
  that is, diffeomorphism and Hamiltonian constraints that do not form
  a Lie algebra. We tackle the problem using the ``uniform
  discretization'' technique. We study the expectation value of the
  master constraint and argue that among the states that minimize the
  master constraint is one that incorporates the usual Fock vacuum for
  the matter content of the theory.
\end{abstract}

\maketitle
\section{Introduction}
Loop quantum gravity is being explored in model situations of
increasing complexity. There has been steady advance in treating
homogeneous cosmologies \cite{lqc}, an area of activity that has come
to be known as loop quantum cosmology. There has also been progress in
spherical symmetry in vacuum \cite{nosotros}. However, in all these
cases one did not have to face the ``problem of dynamics'', i.e.
dealing with the non-Lie algebra of constraints of general relativity.
In homogeneous cosmologies there is only one constraint and it
therefore has a trivial algebra. In spherical symmetry special gauges
were chosen that resulted in an Abelian algebra. In this paper we
would like to study spherically symmetric gravity coupled to a
spherically symmetric scalar field using loop quantum gravity
techniques. It is not known in this situation how to formulate the
problem in a way that one ends up with a Lie algebra of constraints. A
total gauge fixing was introduced by Unruh \cite{unruh}, but it leads
to a non-local expression for the Hamiltonian.  Here we will fix
partially the gauge to eliminate the diffeomorphism constraint in
order to simplify things. This still leads to a Hamiltonian constraint
that has a non-Lie Poisson bracket with itself, involving structure
functions. To treat this problem we will use the ``uniform
discretization'' technique \cite{uniform}. We will introduce a
variational technique adapted to the minimization of the master
constraint (in the context of uniform discretizations one should
probably refer to it as ``master operator'' since it only vanishes in
the continuum limit).  In the case that zero is in the kernel of the
master constraint the technique yields the correct physical state in
model situations.

The inclusion of scalar fields in spherical symmetry opens a rich set
of possibilities to be studied including the formation of black holes,
critical collapse, the emergence of Hawking radiation, among others.
Here we will have much more modest goals: to see how the complete
theory approximates the vacuum state of the scalar field living on a
flat space-time.  An outstanding problem in a full quantum gravity
treatment involving matter fields is the emergence of a vacuum state
for the fields and what relation it may have to the ordinary Fock
vacuum of quantum field theory in curved space-time.  We will apply
the variational technique in the case of spherically symmetric gravity
coupled to a scalar field and show that it yields a vacuum state that
is closely related to the Fock one.

The organization of this paper is as follows: in section II we review
the classical theory. In section III we discuss the quantization of 
a spherical scalar field in a classical flat space-time in order to have
something to compare with the full case. In section IV we study the 
full quantization of gravity and the scalar field, using a variational
technique to minimize the master constraint. We end with a discussion.

\section{Spherically symmetric gravity with a scalar field: the classical theory}

Spherically symmetric gravity with the Ashtekar new variables has been studied
in detail in \cite{spherical} and \cite{boswi}. 
Here we present only a brief summary.
One assumes that the topology of the spatial manifold is of the form
$\Sigma=R^+\times S^2$. We will choose a radial coordinate $x$ and
study the theory in the range $[0,\infty]$. 
The invariant connection can be written 
as,
\begin{eqnarray}
  A &=& A_x(x) \Lambda_3 dx + 
\left(A_1(x) \Lambda_1+ A_2(x) \Lambda_2\right)d\theta\\&& +
\left(\left(A_1(x) \Lambda_2- A_2(x) \Lambda_1\right)\sin \theta +
\Lambda_3 \cos \theta\right) d\varphi,\nonumber
\end{eqnarray}
where $A_x, A_1$ and $A_2$ are real arbitrary functions on $R^+$,
the $\Lambda_I$ are generators of $su(2)$, for instance $\Lambda_I = 
-i\sigma_I/2$ where $\sigma_I$ are the Pauli matrices or
rigid rotations thereof. The invariant triad takes the form,
\begin{eqnarray}
  E &=& E^x(x) \Lambda_3 \sin \theta {\partial \over \partial x} + 
\left(E^1(x) \Lambda_1 + E^2(x) \Lambda_2\right) \sin \theta {\partial \over 
\partial \theta} \nonumber\\&&
+
\left(E^1(x) \Lambda_2 - E^2(x) \Lambda_1\right) {\partial \over 
\partial \varphi},
\end{eqnarray}
where again, $E^x, E^1$ and $E^2$ are functions on  $R^+$. 

As discussed in our recent paper\cite{spherical} and originally by
Bojowald and Swiderski\cite{boswi}, it is best to make several changes
of variables to simplify things and improve asymptotic behaviors. It
is also useful to gauge fix the diffeomorphism constraint to simplify
the model as much as possible.  It would be too lengthy and not
particularly useful to go through all the steps here. It suffices to
notice that one is left with two pairs of canonical variables
$E^\varphi$, ${K}_\varphi$ and $E^x$, $K_x$, and that they are related
to the traditional canonical variables in spherical symmetry
$ds^2=\Lambda^2 dx^2+R^2 d\Omega^2$ by
$\Lambda=E^\varphi/\sqrt{|E^x|}$, $P_\Lambda= -\sqrt{|E^x|}K_\varphi$,
$R=\sqrt{|E^x|}$ and $P_R=-2\sqrt{|E^x|} K_x -E^\varphi
K_\varphi/\sqrt{|E^x|}$ where 
$P_\Lambda$ is the momentum canonically conjugate to $\Lambda$.

In terms of these variables the diffeomorphism and Hamiltonian constraints
for gravity minimally coupled to a massless scalar field are \cite{husain},
\begin{eqnarray}
C_r&=& (|E^x|)' K_x -E^\varphi (K_\varphi)' -P_\phi \phi'\\
H&=& \frac{1}{G}\left[-\frac{E^\varphi}{2\sqrt{|E^x|}} - 2 K_\varphi \sqrt{|E^x|} K_x  
-\frac{E^\varphi K_\varphi^2}{2 \sqrt{|E^x|}}+\frac{\left((|E^x|)'\right)^2}{8\sqrt{|E^x|}E^\varphi}\right.\nonumber\\
&&\left.-\frac{\sqrt{|E^x|}(|E^x|)' (E^\varphi)'}{2 (E^\varphi)^2} +
\frac{\sqrt{|E^x|} (E^x)''\text{sgn}(E^x)}{2 E^\varphi}\right] + \frac{P_\phi^2}{2 \sqrt{|E^x|}E^\varphi}
+\frac{(|E^x|)^{3/2} (\phi')^2}{2 E^\varphi}
\end{eqnarray}
and since the variables are gauge invariant there is no Gauss law. 
We have taken the Immirzi parameter equal to one.
We now proceed to partially fix the gauge by choosing $E^x=x^2$ ($R=x$ in terms of the metric 
variables). One can solve the diffeomorphism constraint for $K_x$, 
\begin{equation}
K_x = \frac{E^\varphi (K_\varphi)'+P_\phi \phi'}{2 x},
\end{equation}
which yields the Hamiltonian constraint for the partially gauge fixed model as,
\begin{eqnarray}
H&=& \frac{1}{G} \left[-\frac{E^\varphi}{2x} -\frac{E^\varphi K_\varphi^2}{2x} + \frac{3 x}{2 E^\varphi}
-\frac{x^2 (E^\varphi)'}{(E^\varphi)^2}-E^\varphi K_\varphi (K_\varphi)'\right] \nonumber\\
&&+\frac{P_\phi^2}{2 x E^\varphi}
+\frac{x^3 (\phi')^2}{2 E^\varphi} -K_\varphi P_\phi \phi'.
\end{eqnarray}

We now rescale the Lagrange multiplier $N_{\rm old}={N}_{\rm new}G (E^x)'/E^\varphi$, the rescaled Hamiltonian 
constraint is,
\begin{equation}\label{ham}
H= H_{\rm vac}+2 G\,H_{\rm matt}
\end{equation}
where 
\begin{eqnarray}
H_{\rm vac}&=&\left( -x -x K_\varphi^2+\frac{x^3}{(E^\varphi)^2}\right)'=\partial H_v(x)/\partial x,\\
H_{\rm matt} &=& \frac{P_\phi^2}{2(E^\varphi)^2}+\frac{x^4 (\phi')^2}{2(E^\varphi)^2} - \frac{x K_\varphi P_\phi 
\phi'}{E^\varphi}.
\end{eqnarray}

This form of the Hamiltonian constraint allows an easy identification
of the required boundary term if one assumes asymptotically flat
conditions.  The total Hamiltonian is given by,
\begin{equation}
H_T=\int_0^{x^+}dx N(x)(H_{\rm vac}(x)+2 G\,H_{\rm matt}(x))+ H_B
\end{equation}
where $N(x)$ is the rescaled lapse $N_{\rm new}$ and $H_B$ is the boundary term at the asymptotic region $x^+$. Integrating by parts we get 
\begin{eqnarray}
H_T&=& -\int_0^{x^+}dx \frac{dN(x)}{dx}\left(H_v(x)+2G\int_0^{x}dy H_{\rm matt}(y)\right)+N(x^+)\left(-2GM+2G\int_0^{x^+}dy H_{\rm matt}(y)\right)+H_B\nonumber \\
&=&- \int_0^{x^+}dx \frac{dN(x)}{dx} \left(H_v(x)-2G\int_x^{x^+}dy H_{\rm matt}(y)+2GM\right)-2GM \dot{\tau}.
\end{eqnarray}
The boundary term $H_B=-2GM\dot{\tau}$ has been introduced in order to
ensure that M is a constant and $\tau$ the proper time in the
asymptotic region. This is the standard boundary term in the
spherically symmetric case. $M$ is the space time mass while the
Schwarzschild radius is given by $R_S=2G(M-\int_0^{x^+}dy H_{\rm
  matt}(y)))$.  In the case of a space time with a black hole the
radial coordinate is given by $R=x+R_S$.  $M$ is a Dirac observable.
In the case of weak fields therefore, so is the integral from $0$ to
$\infty$ of $H_{\rm matt}$ that we shall cal $H_M$. Even in presence
of black holes $H_M$ is an observable if the black hole is isolated.
We will treat $H_M$ as an energy in order to define the vacuum and the
excited states of the theory in the case of interest in this paper,
weak fields without the presence of black holes.

\section{Quantization of the matter field on a fixed flat background}

Since we wish to understand in which way loop quantum gravity recovers
results from ordinary quantum field theory in curved spacetime, we
would like to outline some of those results for later comparison.  If
the space-time is flat it is convenient to fix the gauge $K_\varphi=0$
to obtain explicitly the background metric in the usual spherical
coordinates. In this case one solves $H_{\rm vac}=0$ one gets that
$E^\varphi=x$. Solving the evolution equation yields the Lagrange
multiplier and one recovers the full flat space-time metric. The
matter portion of the Hamiltonian constraint becomes,
\begin{equation}
H_{\rm matt} = \frac{P_\phi^2}{2 x^2} +\frac{x^2 (\phi')^2}{2}.
\end{equation}
The evolution equation obtained from this Hamiltonian corresponds to
spherical waves,
\begin{equation}
\phi''-\ddot\phi+2 \frac{\phi'}{x}=0.
\end{equation}
This can be solved by separation of variables, 
\begin{equation} 
\phi(x,t)= \int_0^\infty d\omega 
\frac{\left(C(\omega) \exp(-i \omega t)+\bar{C}(\omega) \exp(i \omega t)\right)
\sin(\omega x)}{\sqrt{\pi \omega} x}, 
\end{equation} 
which corresponds to spherical waves that are regular at the origin. From
Hamilton's equation we can get an expression for $P_\phi$,
\begin{equation}
P_\phi(x,t) = \int_0^\infty d\omega 
\frac{\left(-i C(\omega) \omega \exp(-i\omega t)+i\bar{C(\omega)}\omega
\exp(i\omega t)\right) x\sin(\omega x)}{\sqrt{\pi \omega}}.
\end{equation}

{}From the standard commutation relations,
$[\hat{\phi}(x,t),\hat{P}_\phi(y,t)]= i\delta(x-y)$, one gets the
$[\hat{C}(\omega),\hat{\bar{C}}(\omega')]=
\delta(\omega-\omega')$.  One can proceed to define a vacuum state
$\vert 0\rangle$ as the state that is annihilated by $\hat{C}$. If one
evaluates the expectation value of $H_{\rm matt}$ on the vacuum state
one finds that it has an ultraviolet divergence. The usual resolution
of this problem is to introduce a cutoff. It should be noted that when
one treats this problem in loop quantum gravity this type of
divergence does not appear because the well defined objects are
holonomies associated to finite paths. In our treatment this aspect is
lost since we have gauge fixed the radial variable which therefore
becomes a c-number. As we usually proceed when we use the uniform
discretization technique, we regularize the expression by placing it
on a lattice. We will discuss later on the issue of taking the lattice 
spacing to zero.

We will assume that the radial direction is bounded with a spatial extent
$L$ and consists of discrete points $x_i$ separated by a coordinate
distance $\epsilon$, and in particular we take $x_i$ as $\epsilon$ times
an integer. We reinterpret the integrals as sums, Dirac deltas as Kronecker
deltas, functional derivatives as partial derivatives, and partial derivatives
in the radial directions as finite differences. Specifically \cite{zako}
\begin{eqnarray}
\int dx &\to& \epsilon \sum_{x}\\
\delta(x-y)&\to& \frac{\delta_{x,y}}{\epsilon}\\
\frac{\delta}{\delta \phi(x)}&\to&\frac{1}{\epsilon}\frac{\partial}{\partial \phi}\\
\phi(x)' &\to& \frac{\phi(x_{i+1})-\phi(x_i)}{\epsilon}\\
(\omega)^2 &\to&\frac{\sum_i \left(2-2\cos(\epsilon \omega_i)\right)}{\epsilon^2}
\end{eqnarray}
If the spatial direction is discrete, the associated momentum space is
bounded with extent $2\pi/\epsilon$. To the first nontrivial order in
in epsilon, all formulae involving momenta $\omega$ are unchanged
except that momentum integrals are now sums over a
momentum space of finite extent.

The expectation value of $\hat{H}_{\rm matt}$ can be computed replacing
the quantum version of the expressions given above for $\phi(x,t)$
and $P_\phi(x,t)$ in $\hat{H}_{\rm matt}$. Computing the expectation
value on the vacuum state one 
is only left with contributions proportional to $\hat{C}\hat{\bar{C}}$. 
On the lattice the result may be
approximated in the limit of large $L$ by the integral,
\begin{equation}
\langle 0\vert \hat{H}_{\rm matt}(x)\vert 0\rangle =
\int_0^{2\pi/\epsilon} d\omega
\frac{\omega^2 x^2-2 x \omega \cos(\omega x) \sin(\omega x) +\sin^2(\omega x)}{2x^2\pi \omega}.
\end{equation}
The integral can be computed in closed form in terms of integral
cosine functions. It is more useful to give an approximation for its value
as an expansion in $\epsilon$,
\begin{equation}
\langle 0\vert \hat{H}_{\rm matt}(x)\vert 0\rangle =
\frac{\pi}{\epsilon^2} -\frac{\sin^2(2\pi x/\epsilon)}{\pi x^2}
+\frac{\ln(x/\epsilon)}{4 x^2 \pi} +O(\epsilon^0).
\end{equation}
The leading order in  the energy density expansion is $\pi/\epsilon^2$ which
has the correct dimensions for an energy density in one spatial
dimension, since we are only considering the radial mode of the scalar
field.

As in four dimensions, the energy of the vacuum gives rise to a
cosmological constant if one allows the field to back-react on
gravity. The nature of this constant is different, however in two
dimensions \cite{thomi}. First of all, notice that if one had started from four
dimensional gravity with a cosmological constant and imposed spherical
symmetry, one can view the model as a $1+1$ dimensional theory with a
dilaton with a mass given by the four dimensional cosmological
constant.  That is, it does not produce a term that behaves like a
cosmological constant in $1+1$ dimensions. The vacuum energy, by
contrast produces a constant term in the Hamiltonian constraint. Second, notice
that even in vacuum $H_{\rm vac}$ already has a constant term in
it. So the energy of the vacuum essentially operates as a rescaling of
that constant term, which in turn can be absorbed by a rescaling of
the radial coordinate. In four dimensions, if one chooses a Planck
scale cutoff it implies that the radius of curvature of space-time
becomes of the order of Planck length, which is clearly unphysical. In
spherical symmetry the presence of the constant can be reabsorbed in a
redefinition of the coordinates.  This redefinition however, has
consequences when one wishes to reinterpret the model as an
approximation to a four dimensional space-time. The redefinition of
the radial coordinate implies that the spheres do not have $4\pi R^2$
area anymore. The four dimensional universe modeled contains a
topological defect, a ``global texture'' \cite{turok}. Notice that this 
immediately precludes taking the lattice spacing to zero, since already
when the lattice spacing is of the order of $\ell_{\rm Planck}$ one will
have a solid deficit angle that exceeds $4\pi$ and does not allow
to interpret the model as a four dimensional space-time.

There are two avenues to handle the situation: either one rescales the
radial variable and accepts that the model approximates four
dimensional space-times with (large) topological defects, or one can
modify the two dimensional model by adding a constant to the
Hamiltonian constraint (a cosmological constant in $1+1$ dimensional
gravity). Such a model will not stem from a dimensional reduction of
four dimensional gravity, but upon quantization will turn out to
approximate four dimensional spherical gravity around a flat
background without a topological defect.

We will take the first point of view and write the Hamiltonian 
constraint as,
$
H= H_{\rm vac}+G\,H_{\rm matt},
$
where 
\begin{eqnarray}\label{21}
H_{\rm vac}&=&\left( -x(1-2\Lambda) -x K_\varphi^2+\frac{x^3}{(E^\varphi)^2}\right)',\\
H_{\rm matt} &=& \frac{P_\phi^2}{(E^\varphi)^2}
+\frac{x^4 (\phi')^2}{(E^\varphi)^2} - 2 \frac{x K_\varphi P_\phi 
\phi'}{E^\varphi}- \rho_{\rm vac},\label{22}
\end{eqnarray}
where $\Lambda= \frac{G}{2} \rho_{\rm vac}$ and $\rho_{\rm
vac}$ is the vacuum energy density. We choose $\hbar=c=1$ units. This rewriting of the constraint has the
property that the expectation value of $H_{\rm matt}$ will be zero in
the vacuum.

\section{Full quantization of the model}

We would like to write the master constraint based on the Hamiltonian
constraint of the model we introduced in the last section. Although
the discrete Hamiltonian constraint fails to close a first
class algebra, we have showed in \cite{difeos} that with the uniform
discretization technique one can consistently treat the problem by
minimizing the resulting master constraint. To write the master
constraint at a quantum level we will polymerize the expression of the
gravitational part of the constraint.  We will not use a polymer
representation in the scalar sector for simplicity and because we want
to make contact with the usual treatments based on a Fock
quantization. It is known that the Fock quantization for fields can be
recovered from the polymer quantization
\cite{thiemannsahlmann,ashtekarshadow}.

\subsection{Variational technique to study the expectation value of the master constraint}

Here we will introduce a variational technique to minimize the master
constraint. The technique is general, it is not restricted to the
model we study in this paper.  We start by considering a fiducial
Hilbert space ${\cal H}_{\rm aux}$ in which the master constraint is a
well defined self-adjoint operator.  We will then use a variational
technique to find approximations to the minimum value of the
expectation value of the master constraint within this space. In many
cases of interest, the minimum expectation value will not be zero, but
will be small (the master constraint has units of action squared, so
normally one would require it to be much smaller than $\hbar^2$, in
order to have a good approximation of the physical space, in our units
that translates into much smaller than one). As we will see in the
examples, the resulting quantum theory will therefore not reproduce
exactly the symmetries of the continuum theory but it will approximate
them, even at the quantum level. We will see that if zero is in the
spectrum of the operator the corresponding eigenstates in many cases
will be distributional with respect to the fiducial space we are
considering.

To implement the variational method, we consider trial states in
${\cal H}_{\rm aux}$ that are Gaussians centered around the classical
solution of the model of interest in phase space.  That means that as
functions of ${\cal H}_{\rm aux}$ these will generically be Gaussians
times phase factors such that the resulting state is centered around
the classical solution in both configuration variables and momenta.
The states are parameterized by the values of the standard deviations
of the Gaussians in either configuration or momentum space.  A caveat
is that in gauge theories one may choose to work with a classical
solution that is not in a completely determined gauge.  Such a
solution will be a trajectory in phase space. Such a trajectory will
determine some of the canonical variables as functions of others,
which will remain free. In that case one has to allow such variables
to be free in the trial solution by considering Gaussians centered
around a value that is a free parameter.  If one chooses to work with
a classical solution in a completely specified gauge one just
considers Gaussians around the point in phase space represented by the
classical solution of interest and extremizes the expectation value of
the master constraint with respect to the standard deviations of the
Gaussians. It can
happen that the extremum occurs as a limit in the parameter space in
which case the resulting state does not belong in ${\cal H}_{\rm aux}$
but in its dual (after a suitable rescaling, it becomes a
distribution).

Before attacking the problem of interest, it is useful to see the
technique we just described in action in a couple of simple examples. 
The first example
we choose is a system with two degrees of freedom $q_1,p_1$ and
$q_2,p_2$, and two constraints $p_1=0$ and $p_2=0$. The total
Hamiltonian for the system is $H_T=N_1 p_1+N_2 p_2$ with $N_{1,2}$
Lagrange multipliers. The states annihilated by the constraints are
trivial and given by the distribution $\delta(p_1)\delta(p_2)$.  We
fix a gauge $q_1-q_2=0$. Fixing the gauge is not needed in a simple
model like this, but may be a necessity to simplify things in more
complicated models. So we will choose a gauge fixing here to show that
in the end the process loses all information about the gauge fixing
and recovers the correct physical state. This requires fixing the
Lagrange multipliers so there is only one ($N$) left and the total
Hamiltonian becomes $H_T= N(p_1+p_2)$. The conjugate variable to the
gauge fixing, $p_1-p_2$ is strongly zero. We start with a two
parameter family of states in ${\cal H}_{\rm aux}$ choosing as
configuration variables $q_1-q_2$ and $p_1+p_2$,
\begin{equation}
\psi_{\sigma_\pm,\beta} = \frac{1}{\sqrt{\pi \sqrt{\sigma_+\sigma_-}}}
\exp\left(-\frac{\left(q_1-q_2\right)^2}{2 \sigma_-}\right)
\exp\left(-\frac{\left(p_1+p_2\right)^2}{2 \sigma_+}\right)
\exp\left(i\beta\left(p_1+p_2\right)\right),\label{26}
\end{equation}
with $\beta$ an arbitrary parameter associated with the fact that the
variable $q_1+q_2$ is pure gauge.  One could choose to work in a
completely gauge fixed solution in which $q_1+q_2$ is zero, in that
case there is no need to introduce the parameter $\beta$.  The choice
of this family of states is based on the fact that they describe
wave-packets centered around the classical solutions of the
constraints, $q_1-q_2=0$, $p_1-p_2=0$ and $p_1+p_2=0$.  We now define
the master constraint ${\mathbb H}=p_1^2+p_2^2$ and act on this space
of states. The expectation value is,
\begin{equation}
\langle \psi_{\sigma_\pm,\beta}\vert {\mathbb H} \vert  
\psi_{\sigma_\pm,\beta}\rangle = 
\frac{1}{4\sigma_-} +
\frac{1}{4} \sigma_+
\end{equation}
where $\sqrt(\sigma_\pm)$ are the standard deviations of the Gaussians, $\sigma_\pm $ taken to be positive. One therefore sees that the expectation
value cannot be zero for any finite value of the $\sigma$'s. However,
if one takes $\sigma_-=\frac{1}{2\epsilon^2}$ and $\sigma_+=2\epsilon^2$ then
in the limit $\epsilon\to 0$, $<{\mathbb H}>=O(\epsilon^2)$. The states $\vert
\psi_\epsilon\rangle$ become,
\begin{equation}
\langle q_1-q_2,p_1+p_2\vert \psi_\epsilon\rangle = \frac{1}{\sqrt{\pi}}
\exp\left(-\left(q_1-q_2\right)^2\epsilon^2\right)
\exp\left(-\frac{\left(p_1+p_2\right)^2}{4\epsilon^2}\right)
\exp\left(i\beta\left(p_1+p_2\right)\right),
\end{equation}
And their Fourier transform
\begin{equation}
\langle p_1-p_2,p_1+p_2\vert \psi_\epsilon\rangle = \frac{1}{\epsilon\sqrt{2\pi}}
\exp\left(-\frac{\left(p_1-p_2\right)^2}{4\epsilon^2}\right)
\exp\left(-\frac{\left(p_1+p_2\right)^2}{4\epsilon^2}\right)
\exp\left(i\beta\left(p_1+p_2\right)\right),
\end{equation}

These states are normalized in ${\cal H}_{\rm aux}$ but they vanish
(in the sense of distributions) in the limit $\epsilon\to 0$. They
need to be rescaled in order to end up with well defined distribution
on some suitable subspace of ${\cal H}_{\rm aux}$.

So the physical states would be
\begin{equation} 
\langle p_1-p_2,p_1+p_2|\psi\rangle_{\rm ph} \equiv \lim_{\epsilon\to 0} 
\frac{1}{\sqrt{2\pi}\epsilon}
\langle p_1-p_2,p_1+p_2
\vert
\psi_\epsilon\rangle= 2\delta(p_1+p_2)\delta(p_1-p_2)=\delta(p_1)\delta(p_2)
\end{equation} 
Notice that the parameter $\beta$ is free at the end of the process since
it corresponds to the value of a variable that is pure gauge in this model.

There is an additional element that the above example does not capture
and we would like to discuss. When we apply this technique in
situations of interest, we will be discretizing the theories we
analyze. Usually, discretization turns first class constraints into
second class ones. The uniform discretization procedure tells us that
we do not need to concern ourselves with the second class nature of
the constraints (for a discussion see \cite{difeos}). We can still
consider the master constraint and seek the minimization of its
eigenvalues, but the presence of second class constraints in the
discrete theory usually implies that the minimum eigenvalue of the
master constraint will not be zero. The best one can hope for is that
it will be small and the resulting quantum theory will approximate the
symmetries of the theory one started with. This is a point of view
that has been held as natural for some time in the context of quantum
gravity, where one expects that some level of fundamental discreteness
will emerge.  We would like to illustrate this with a modification of
the previous example. Instead of taking $p_1=0$ and $p_2=0$ as the
constraints we will take $p_1+\alpha q_2=0$ and $p_2=0$ with $\alpha$
a small parameter (in realistic theories the small parameter is
related to the lattice spacing in the discretization). We will still
take the same set of $\psi_{\sigma_\pm,\beta}$ as before, that is, for
the trial solution we have chosen Gaussians centered around classical
solutions of the gauge theory where the anomalous term vanishes.  We
do this because one usually knows solutions to the continuum theory
one wishes to approximate (e.g. flat space or the Schwarzschild
solution in the case of gravity) whereas the discrete theories have
complicated solutions that usually cannot be treated in analytic form.
The master constraint now becomes,
\begin{equation}
{\mathbb H} = p_1^2+p_2^2+2\alpha p_1 q_2 +\alpha^2 q_2^2,
\end{equation}
and using the same ansatz (\ref{26}) for the states one finds that
\begin{equation}
\langle \psi_{\sigma_\pm,\beta}\vert {\mathbb H} \vert  
\psi_{\sigma_\pm,\beta}\rangle = 
{\alpha}^{2}{\beta}^{2}+\frac{1}{4\sigma_-} +     \frac{1}{4}\sigma_+ +{\frac {
{\alpha}^{2}}{2\sigma_+}}+\frac{{\alpha}^{2}{\sigma_-}}{2}.
\end{equation}
We would like to identify a limit in the variables $\sigma_\pm$ such that
this quantity vanishes. As was to be expected, this is not possible. 
We can attempt to find values of the parameters $\sigma_\pm$ and
$\beta$ that minimize this expression. The result is $\beta=0$ 
and $\sigma_+=\sqrt{2}{\alpha}$ and $\sigma_-=\frac{{1}}{\sqrt{2}\alpha}$.
which yields $\langle \psi_{\rm min}  \vert
 {\mathbb H} \vert
\psi_{\rm min}
\rangle = {\sqrt{2}\alpha}$. The state is,
\begin{equation}
\langle p_1,p_2\vert\psi_{\rm min}\rangle=
\exp\left(
-\frac{\left({p_1}^{2}+{p_2}^{2} \right)\sqrt {2}}
{\alpha}\right)
\sqrt{\frac {\sqrt{2}}{{\alpha\pi}}}.
\end{equation}

It is interesting to compare this state and the corresponding expectation value of ${\mathbb H}$ obtained from our variational technique with the exact minimum of this model. 
A naive analysis would tell us that the minimum corresponds to an
exact eigenstate with zero eigenvalue for ${\mathbb H}$. However, that
solution is not well behaved.  It is known that one can find solutions
of the master constraint that do not solve the constraints if one does
not impose regularity in the solutions found \cite{thiemannmc2}.  The
master constraint is an operator in the Hilbert space and one can
analyze its spectral resolution. The spurious solutions do not belong
in the spectral resolution of the master constraint. In this case one
can solve exactly the eigenvalue problem ${\mathbb H} \vert
\psi\rangle =E \vert \psi\rangle$. The solutions with minimum eigenvalue
are of the form
$\delta(p_1) \psi_0(p_2)$ where $\psi_0(p_2)$ is the fundamental state of 
the Hamiltonian of a harmonic oscillator in the momentum
representation. The minimum eigenvalue for such exact solution is
$\alpha$ (compare with the variational one in which the eigenvalue was
slightly higher $\sqrt{2}\alpha$). It is also interesting to note that
if instead of choosing the gauge $q_1-q_2=0$ we had chosen $q_1=0$ and
proceeded with the variational technique, one obtains the exact state
directly. This illustrates that the method approximates well the state
of interest in situations where zero is not in the kernel of the
master constraint. The solution that minimizes the master constraint
admits a very simple interpretation that shows that the uniform
discretization of the theory with the anomalous term $\alpha$ small but
non-vanishing, approximately reproduces the invariances of the theory
with first class constraints $p_1=P_2=0$. In fact $q_1$ and $q_2$ are
gauge variables and the physical space is independent of these
variables. The physical state is constant in $q_1$ and $q_2$. For a
small but non vanishing alpha the physical states are independent of
$q_1$ and weakly dependent on $q_2$. A final comment is that in
this case the parameter $\beta$, which was not determined in the 
case with first class constraints, gets determined here. That is, in
the case where $\beta$ was associated with an exact gauge symmetry,
the minimization of the master constraint was insensitive to the value 
of $\beta$. In the case where the constraints are second class and 
we do not get zero as minimum of the master constraint there is
some dependence on $\beta$, but it is weak, since the term in the
master constraint is $\beta^2 \alpha^2$ and $\alpha$ is small (in the
quantum state one has approximately $\delta(p) \exp(i p \beta)$). 
The theory where one does not exactly annihilate the master constraint
only has approximate gauge symmetries and therefore has slightly 
``preferred'' gauges from the point of view of minimizing the master
constraint.

\subsection{The discrete master constraint}

Let us now consider the complete Hamiltonian constraint. We wish to
discretize it and to polymerize the gravitational variables.  The
Hamiltonian gets rescaled in the discretization $H(x_i) \to
H(i)/\epsilon$,. We also rescale the expression multiplying the
continuum Hamiltonian constraint times $G$. 
The resulting discrete expression is,
\begin{eqnarray}
H(i) &=& -(1-2\Lambda)\epsilon \label{33}
- x(i+1)\frac{\sin^2\left(\rho K_\varphi(i+1)\right)}{\rho^2} 
+ x(i)\frac{\sin^2\left(\rho K_\varphi(i)\right)}{\rho^2} 
+ \frac{x(i+1)^3 \epsilon^2}{(E^\varphi(i+1))^2}
-  \frac{x(i)^3 \epsilon^2}{(E^\varphi(i))^2}\\
&&
+G\left(\epsilon\frac{(P^\varphi(i))^2}{(E^\varphi(i))^2}
+\epsilon\frac{x(i)^4 \left(\phi(i+1)-\phi(i)\right)^2}{(E^\varphi(i))^2}
-2 x(i)\frac{\sin\left(\rho K_\varphi(i)\right)}{E^\varphi(i)\rho} 
{\left(\phi(i+1)-\phi(i)\right)} P^\phi(i)- \rho_{\rm vac}\epsilon 
\right)\nonumber.
\end{eqnarray}

We need to construct the master constraint. Since the Hamiltonian 
is a density of weight one, we define the master constraint associated
with the Hamiltonian constraint in  the full theory as,
\begin{equation}
{\mathbb H} =\frac{1}{2} \int dx \frac{H(x)^2}{\sqrt{g}}\ell_{\rm P},
\end{equation}
or, in terms of the variables of the model, up to a constant factor,
\begin{equation}
{\mathbb H} = \frac{1}{2}\int dx \frac{H(x)^2}{\left(E^\varphi\right)
\sqrt{E^x}}\ell_{\rm P},
\end{equation}
and in the discretized theory ${\mathbb H}^\epsilon = 
\sum_i {\mathbb H}(i)$ with
\begin{equation}
{\mathbb H}(i) = \frac{1}{2} 
\frac{{H(i)^2}\ell_{\rm P}}{\sqrt{E^x(i)}E^\varphi(i)} .\label{36}
\end{equation}
The constant $\ell_{\rm P}$ must be introduced so that $\mathbb{H}$ is
dimensionless with $\hbar=c=1$, one could use $\sqrt{G}$ instead of
it.  It is convenient to rescale the Hamiltonian constraint by
$\sqrt{E^\varphi/(E^x)'}$.  This does not change the density
weight. If one does not rescale things it turns out ${\mathbb H}$ is
proportional to $1/E^\varphi$.  In the polymer representation this
implies that the vacuum is the ``zero loop'' state, which is
degenerate (it corresponds to zero volume space-times). To eliminate
this unphysical possibility one exploits the fact that the Hamiltonian
constraint is defined up to a factor given by a scalar function of the
canonical variables without changing the first class nature of the
classical constraint algebra. The rescaling factor in the discrete
theory after the gauge fixing is
$\sqrt{E^\varphi(i)/(2x(i)\epsilon)}$. So (\ref{33}) has to be multiplied
times that factor when constructing the master constraint (\ref{36}).

Let us focus on the matter portion of the Hamiltonian, we will write it as,
\begin{equation}\label{matterparthamiltonian}
H_{\rm matt}(i) = 
\frac{H_{\rm matt}^{(1)}(i)}{(E^\varphi)^2(i)}
+
\frac{H_{\rm matt}^{(2)}(i)\sin\left(\rho K_\varphi(i)\right)}{\rho E^\varphi(i)}
-H_{\rm matt}^{(3)}(i).
\end{equation}

The master constraint can be written as,
\begin{eqnarray}\label{masterconstraint}
  {\mathbb H}(i) &=& \ell_{\rm P}\left[c_{11}(i) \left(H_{\rm matt}^{(1)}(i)  \right)^2
+c_{22}(i) \left(H_{\rm matt}^{(2)}(i)  \right)^2 \right.\\
&&+ c_{1}(i) H_{\rm matt}^{(1)}(i)  
+ c_{2}(i) H_{\rm matt}^{(2)}(i)  
+ c_{33}(i) \left(H_{\rm matt}^{(3)}(i)\right)^2
+ c_{3}(i) H_{\rm matt}^{(1)}(i)  \nonumber\\
&&\left.+ c_{12}(i) H_{\rm matt}^{(1)}(i) H_{\rm matt}^{(2)}(i)
+ c_{13}(i) H_{\rm matt}^{(1)}(i) H_{\rm matt}^{(3)}(i)
+ c_{23}(i) H_{\rm matt}^{(2)}(i) H_{\rm matt}^{(3)}(i)\nonumber
+ c_{00}(i)\right],
\end{eqnarray}
where,
\begin{eqnarray}
H_{\rm matt}^{(1)}(i) &=&\left(
\epsilon \left(P^\varphi(i)\right)^2
+\epsilon x(i)^4 \left(\phi(i+1)-\phi(i)\right)^2\right)\ell^2_P \\   
H_{\rm matt}^{(2)}(i) &=&\left(
-2x(i)\left(\phi(i+1)-\phi(i)\right) P^\varphi(i)\right)\ell^2_P
\\   
H_{\rm matt}^{(3)}(i) &=&2\rho_{\rm vac}\epsilon \ell^2_P.
\end{eqnarray}
To economize space, we will not give the classical expressions
for the coefficients, since they can be readily obtained from the
quantum expressions.

In order to quantize the master constraint we need to choose a factor
ordering. The expression of the master constraint is a sum of symmetric
operators consisting of polynomials in $\hat{E}^\varphi$ and $\sin(\rho
\hat{K}_\varphi)$, $\hat{P}^\phi$ and $\hat{\phi}$.  We choose a
factor ordering with the factors of $\hat{E}^\varphi$ are distributed
symmetrically to the right and the left of the factors of $\sin(\rho
\hat{K}_\varphi)$.  For the factors $\hat{P}^\phi$ and $\hat{\phi}$ we
follow a similar strategy, putting the $\hat{P}^\phi$ symmetrically to
the left and to the right of $\hat{\phi}$'s. The coefficients in the above expression
of the master constraint with this factor ordering are,
\begin{eqnarray}\label{44}
\hat{c}_{11}(i)&=&\frac{1}{4x(i)^{2}\epsilon\hat{E}^{\varphi}(i)^4},\\
\hat{c}_{12}(i)&=&\frac{1}{2x(i)^{2}\rho\epsilon}\frac{1}{\hat{E}^{\varphi}(i)^{3/2}}\sin(\rho K_{\varphi}(i))\frac{1}{\hat{E}^{\varphi}(i)^{3/2}},\\
\hat{c}_{13}(i)&=&-\frac{1}{2x(i)^{2}\epsilon}\frac{1}{\hat{E}^{\varphi}(i)^2},\\
\hat{c}_{22}(i)&=&\frac{1}{8x(i)^{2}\rho^2\epsilon}\left(\frac{1}{\hat{E}^{\varphi}(i)^2}-\frac{1}{\hat{E}^{\varphi}(i)}\cos(2\rho K_{\varphi}(i)) \frac{1}{\hat{E}^{\varphi}(i)}\right),\\
\hat{c}_{23}(i)&=&-\frac{1}{2x(i)^{2}\rho\epsilon}\frac{1}{\sqrt{\hat{E}^{\varphi}(i)}}\sin(\rho K_{\varphi}(i))\frac{1}{\sqrt{\hat{E}^{\varphi}(i)}},\\
\hat{c}_{33}(i)&=&\frac{1}{4x(i)^{2}\epsilon},\\
\hat{c}_{1}(i)&=&-\frac{x(i)\epsilon}{2\hat{E}^{\varphi}(i)^4}
+\frac{1}{4\epsilon x(i)^2\hat{E}^{\varphi}(i)}\left(-2\epsilon(1-2\Lambda) + \frac{x(i+1)\cos(2\rho K_{\varphi}(i))}{\rho^2}-\frac{x(i+1)}{\rho^2}\right)\frac{1}{\hat{E}^\varphi(i)} \nonumber\\
&&-\frac{1}{\hat{E}^{\varphi}(i)} \frac{\cos(2\rho K_{\varphi}(i))}{4x(i)\rho^2\epsilon} \frac{1}{\hat{E}^{\varphi}(i)}+ \frac{1}{4x(i)\rho^2\epsilon\hat{E}^{\varphi}(i)^2} + \frac{\epsilon x(i+1)^{3}}{2 x(i)^{2}}\frac{1}{\left(\hat{E}^{\varphi}(i)\hat{E}^{\varphi}(i+1)\right)^{2}},\\
\hat{c}_{2}(i)&=&\left[
-\frac{1}{2\rho x(i)^2}\left(1-2\Lambda\right)
+\frac{x(i+1)}{4\rho^3x(i)^2\epsilon}
\left(\cos(2\rho K_\varphi(i+1))-1\right)+\frac{3}{8\rho^3x(i)\epsilon}
+\frac{\epsilon x(i+1)^3}{2 \rho x(i)^2\hat{E}^\varphi(i+1)^2}
\right]\times\nonumber\\
&&\times
\frac{1}{\sqrt{\hat{E}^{\varphi}(i)}}
\sin(\rho K_\varphi(i))
\frac{1}{\sqrt{\hat{E}^{\varphi}(i)}}
-\frac{1}{8\rho^3x(i)\epsilon}
\frac{1}{\sqrt{\hat{E}^{\varphi}(i)}}
\sin(3\rho K_\varphi(i))
\frac{1}{\sqrt{\hat{E}^{\varphi}(i)}}\nonumber\\
&&
-\frac{x(i)\epsilon}{2\rho}
\frac{1}{{\hat{E}^{\varphi}(i)^{3/2}}}
\sin(\rho K_\varphi(i))
\frac{1}{{\hat{E}^{\varphi}(i)^{3/2}}}
\\
\hat{c}_{3}(i)&=&\frac{1}{2x(i)^2}(1-2\Lambda)
+\frac{x(i+1)}{4x(i)^2\epsilon\rho^2}
(1-\cos(2\rho K_{\varphi}(i+1)))
-\frac{1}{4x(i)\epsilon\rho^2}(1-\cos(2\rho K_{\varphi}(i)))
\nonumber\\
&&+\frac{x(i)\epsilon}{2\hat{E}^{\varphi}(i)^2}-\frac{\epsilon x(i+1)^3}{2x(i)^2 \hat{E}^{\varphi}(i+1)^2},\\
\hat{c}_{00}(i)&=&\frac{1}{32\epsilon\rho^4} \left(3-4\cos(2\rho K_{\varphi}(i))+\cos(4\rho K_{\varphi}(i))\right)\nonumber\\
&&+\frac{\epsilon}{4x(i)^2}+\frac{x(i+1)}{4x(i)^2\rho^2}(1-\cos(2\rho K_{\varphi}(i+1)))\nonumber\\
&&-\frac{x(i+1)}{8x(i)\epsilon\rho^4} \left(1-\cos(2\rho K_{\varphi}(i))-\cos(2\rho K_{\varphi}(i+1))+\cos(2\rho K_{\varphi}(i)) \cos(2\rho K_{\varphi}(i+1) )\right)\nonumber\\
&&+\frac{x(i+1)^2}{32\epsilon\rho^4 x(i)^2}\left(3+\cos(4\rho K_{\varphi}(i+1))-4\cos(2\rho K_{\varphi}(i+1)) \right)\nonumber\\
&&-\frac{\Lambda x(i+1)}{2x(i)^2\rho^2}\left(1-\cos(2\rho K_{\varphi}(i+1)) \right)\nonumber\\
&&-\frac{1}{4x(i)\rho^2}(1-2\Lambda)(1-\cos(2\rho K_{\varphi}(i)))
-\frac{\epsilon\Lambda}{x(i)^2}(1-\Lambda)\nonumber\\
&&+\frac{\epsilon x(i+1)^3}{4x(i)\rho^2}\left(\frac{1}{\hat{E}^{\varphi}(i+1)^2}-\frac{1}{\hat{E}^{\varphi}(i+1)}\cos(2\rho K_{\varphi}(i))\frac{1}{{\hat{E}^{\varphi}(i+1)}} \right)\nonumber\\
&&-\frac{x(i)\epsilon}{4\rho^2}
\left(x(i)\left(\frac{1}{\hat{E}^{\varphi}(i)^2}
- \frac{1}{\hat{E}^{\varphi}(i)} \cos(2\rho K_{\varphi}(i)) 
\frac{1}{\hat{E}^{\varphi}(i)} \right) \right.\nonumber\\
&&\left.
- x(i+1) \left(\frac{1}{\hat{E}^{\varphi}(i)^2} -\frac{1}{\hat{E}^{\varphi}(i)}\cos(2\rho K_{\varphi}(i+1)) \frac{1}{\hat{E}^{\varphi}(i)} \right) \right)\nonumber\\
&&-\frac{\epsilon x(i+1)^4}{4x(i)^2\rho^2}\left(\frac{1}{\hat{E}^{\varphi}(i+1)^2} -\frac{1}{\hat{E}^{\varphi}(i+1)} \cos(2\rho K_{\varphi}(i+1)) \frac{1}{\hat{E}^{\varphi}(i+1)} \right)+\frac{x(i)\epsilon^2}{2\hat{E}^{\varphi}(i)^2}(1-2\Lambda)\nonumber\\
&&-\frac{\epsilon^2 x(i+1)^3}{2x(i)^2 \hat{E}^{\varphi}(i+1)^2}(1-2 \Lambda) + \frac{\epsilon^3 x(i+1)^6}{4x(i)^2\hat{E}^{\varphi}(i+1)^4}-\frac{x(i)\epsilon^3 x(i+1)^3}{2\left(\hat{E}^{\varphi}(i+1)\hat{E}^{\varphi}(i)\right)^2} + \frac{x(i)^4\epsilon^3}{4\hat{E}^{\varphi}(i)^4},\label{53}
\end{eqnarray}
and it should be noted that the coefficients commute with $H_{\rm
matt}^{(1)}, H_{\rm matt}^{(2)}$ and $H_{\rm matt}^{(3)}$ so there are
no ordering issues with them.

\subsection{Construction of the trial states}
Since we are interested in the vacuum solution, that classically
corresponds to vanishing scalar fields, we will therefore ignore
$H_{\rm matt}$ (\ref{22}) and only consider the gravitational part
(\ref{21}) in order to construct the classical solution used to build
the ansatz states for the variational technique,
\begin{equation}
H_{\rm vac}=\left( -x(1-2\Lambda) -x
K_\varphi^2+\frac{x^3}{(E^\varphi)^2}\right)'.
\end{equation}

As we discussed in subsection A  we will choose a definite
gauge to work in. Our choice is $K_\varphi=0$, and this implies
$E^\varphi= x/\sqrt{1-2\Lambda}$.  As we claimed before, the presence
of the cosmological constant rescales the radial variable (recall that
without the constant the solution was $E^\varphi=x$). The resulting
four dimensional space-time will be locally flat with a  solid deficit
angle and described in spherical coordinates.

We construct a polymer representation. As we did in previous papers
\cite{spherical} 
one sets up a lattice of points $j=0\ldots N$ in the radial direction
and writes a ``point holonomy'' for the $K_\varphi$ variable at each
lattice site,
\begin{equation}
T_{\vec{\mu}} = \exp\left(i\sum_j \mu_j K_\varphi(j)\right) =\langle 
K_\varphi\vert\, \vec{\mu}\,\rangle.
\end{equation}
In this expression the quantities $\mu_i$ play the role of the ``loop''
in this one dimensional context. They also are proportional to 
the eigenvalues of the
triad operator $\hat{E}^\varphi(i)$. 
The quantum state we will choose for the variational method will be
centered around the classical solution and therefore  we will choose to have
the variable $\mu_i$ centered at the classical value of
$E^\varphi(i)=\epsilon x_1(i)\equiv\epsilon x(i)/\sqrt{1-2\Lambda}$,
\begin{equation} \label{state}
\langle\, \vec{\mu}\,\vert \psi_{\vec{\sigma}}\rangle = \prod_i \sqrt[4]{\frac{2}{\pi\sigma(i)}}
\exp\left(-\frac{1}{\sigma(i)}
\left(\mu_i-\frac{x_1(i)\epsilon}{\ell_{\rm P}^2}\right)^2
\right)
\end{equation} 
on this state $\langle E^\varphi(i)\rangle =\epsilon x_1(i)$ and
$\langle K_\varphi(i) \rangle =0$. Notice that this type of ansatz in
general will be too restrictive: we have ignored possible correlations
among neighboring points by assuming a Gaussian at each point. This could
potentially be problematic when studying excited states and computing
propagators. We will not attack those problems in this paper so we 
will continue with the restrictive ansatz for the moment being.

We will now compute the expectation value of the matter portion of the
Hamiltonian constraint on the above state. The result will be an
operator acting on the matter fields. We will then construct the
vacuum for the resulting operator. What we are doing is to construct a
quantum field theory living on the geometry given by the expectation
values of the triad and extrinsic curvature on the above state.  We
proceed in this way for expediency since this is our first approach to
the problem. In the future we plan to revisit the problem treating all
the variables in a polymerized representation, both gravitational and
material ones, with the variational technique. Preliminary results
indicate that such an approach is viable. For the matter field one
would start by considering a coherent state centered at zero values
for the field and then will obtain the vacuum as a limit. This would
yield valuable insights into the relation of the usual Fock
quantization with the loop quantum gravity techniques, especially
when one gets to discuss physical elements like the propagators of
fields.

In order to take the expectation value of the matter portion of the
Hamiltonian constraint, 
(\ref{matterparthamiltonian}) on the state (\ref{state})
we need to realize two quantum operators. The first one is,
\begin{eqnarray}
  \frac{1}{\left(\hat{E}^\varphi(i)\right)^2}
\langle\, {\mu(i)}\,\vert \psi_{\sigma(i)}\rangle
 &=& 
\left(\frac{2}{3}\right)^{12}\left| \mu(i) \right|  \left(  \left(  \left|
\mu(i)+\rho \right|  \right) ^{3/4}- \left(  \left| \mu(i)-\rho \right| 
 \right) ^{3/4} \right) ^{12}\sqrt [4]{{\frac {2}{\pi \,
\sigma(i)}}}\exp\left({- \frac{\left( \mu(i)-\frac {\epsilon\,x_1(i)}{\ell_{\rm P}^2}
 \right) ^{2}}{\sigma(i)}}\right),
\end{eqnarray}
where we have considered the action on one of the factors of (\ref{state}).
To derive this expression
we consider $\left(\hat{E}\right)^{-3/2} \hat{E}
\left( \hat{E}\right)^{-3/2}$ and use the realization of $\left(
  \hat{E}\right)^{-3/2}$ that was discussed in the context of 
loop quantum cosmology in \cite{AsPaSi}. The reason we can use the loop
quantum cosmology results is that our Hilbert space is a direct product
of loop quantum cosmology Hilbert spaces each at one of the lattice sites
in the radial direction. With the above result one can compute the 
expectation value,
\begin{equation}\label{expectunosobreE}
\langle\,  \psi_{\vec{\sigma}} \vert
\frac{1}{\left(\hat{E}^\varphi(i)\right)^2}
 \vert \psi_{\vec{\sigma}}\rangle=
{\frac {1-2\,\Lambda}{{\epsilon}^{2}{x(i)}^{2}}}+
\frac{5}{8}\,{\frac {{{\ell_{\rm P}}}^
{4} \left( 1-2\,\Lambda \right) ^{2}{\rho}^{2}}{{\epsilon}^{4}{x(i)}^{4}}
}+\frac{3}{4}\,{\frac {\sigma\,{{\ell_{\rm P}}}^{4} \left( 1-2\,\Lambda \right) ^{2}
}{{\epsilon}^{4}{x(i)}^{4}}}.
\end{equation}
The calculation is done by integrating in $\vec{\mu}$ and the result
is lengthy, here we just show it in the approximation $\epsilon> \ell_{\rm P}$.
The first term is the classical value, the others are quantum corrections,
the first one comes from the polymerization, the second from fluctuations in
$\vec{\mu}$. The second operator we need is the one arising in the second
term of the Hamiltonian,
\begin{equation}
\langle\,  \psi_{\vec{\sigma}} \vert
\frac{1}{\sqrt{\hat{E}^\varphi(i)}}
\frac{\sin(\rho \hat{K}_\varphi(i))}{\rho}
\frac{1}{\sqrt{\hat{E}^\varphi(i)}}
\vert\,  \psi_{\vec{\sigma}} \rangle=0.
\end{equation}
To quickly see why this is zero keep in mind that the state is a
Gaussian centered at $K_\varphi=0$ and the sine is an odd function.
With these results the expectation value of the Hamiltonian (the
``effective Hamiltonian'') is,
\begin{equation}\label{effectiveham}
\hat{H}^{\rm eff}_{\rm matt}=
\langle\,  \psi_{\vec{\sigma}} \vert
\hat{H}_{\rm matt}(x,t)
\vert\,  \psi_{\vec{\sigma}} \rangle=
 \frac{\left( 1-2\,\Lambda \right)\left(\hat{P}^\phi(x,t)\right)^{2}}
{x^2 g(x)^2}
+\frac{x^2 \left(1-2\,\Lambda \right) \left(\hat{\phi}'(x,t)\right)^2}
{g(x)^2}-\rho_{\rm vac}.
\end{equation}
In this equation we have pursued the unusual approach of taking the
continuum limit in the terms that involve derivatives and the terms
that involve the momenta of the scalar field. This simplifies
calculations since we will be dealing with differential equations
rather than difference equations. The idea is that the solutions to
the differential equations, suitably discretized, will be a good 
approximation (at least to $O(\epsilon)$ corrections) 
to the solutions of the difference equations.  In the above expression
the quantity $g(x)$ is given by,
\begin{equation}
  g(x)=1-{\frac {5}{16}}\,{\frac {{{\ell_{\rm P}}}^{4}{\rho}^{2}{(1-2\Lambda)}}{{x}^
{2}{\epsilon}^{2}}}-
\frac{3}{8}\frac{\sigma(x)\,\ell_{\rm P}^{4}(1-2\Lambda)}
{{x}^{2}{\epsilon}^{2}}.
\end{equation}

{}From the effective Hamiltonian we get the ``wave equation'' for the fields
living on the curved semiclassical background,
\begin{equation}
  \frac{2}{x}\,{\frac {\partial\phi(x,t)}{\partial x}}
-\frac{2}{g(x)}\frac{\partial \phi(x,t)}{\partial x} 
\frac{\partial g(x)}{\partial x}
+{\frac {\partial ^{2}\phi  \left( x,t \right) }{\partial x^2}}
-\frac{1}{4}\frac{g(x)^4}{(1-2\Lambda)^2}
\frac{\partial^2 \phi(x,t)}{\partial t^2} =0.
\end{equation}

Since the background is time-independent, positive and negative frequency
modes can be introduced by going to Fourier space in $t$. The resulting
equation can be cast in Sturm--Liouville form as,
\begin{equation}
  \left(2 B(x) \phi'(x,\omega)\right)'+ \frac{\omega^2}{2} \phi(x,\omega) A(x)=0
\end{equation}
where
\begin{eqnarray}
  A(x)&=&{\frac {{x}^{2}}{1-2\,\Lambda}}-\frac{5}{8}\,{\frac {{\ell_{\rm P}}^{4}{\rho}^{2}
}{{\epsilon}^{2}}}-\frac{3}{4}\,{\frac {\sigma\,{\ell_{\rm P}}^{4}}{{\epsilon}^{2}
}}
,\\
B(x) &=& {x}^{2} \left( 1-2\,\Lambda \right) +\frac{5}{8}\,{\frac {{\ell_{\rm P}}^{4}{\rho}
^{2} }{{\epsilon}^{2}}}+\frac{3}{4}\,{\frac {
\sigma\,{\ell_{\rm P}}^{4} }{{\epsilon}^{2
}}}
\end{eqnarray}

The solution to this Sturm--Liouville problem is 
\begin{eqnarray}\label{sturmsol}
 \phi(x,w) &=& 
\frac{1}{x} \sin \left( \frac{\omega\,x}{2\left(1-2\Lambda\right)} 
\right)\\&&-
\frac{1}{3x^3}\, 
\left[
x^2 \omega^2 \cos\left(\frac{\omega x}{2}\right){\rm Si}\left(\omega x\right)
-\frac{x}{2} \omega\cos\left(\frac{\omega x}{2}\right)
-x^2 \omega^2 \sin\left(\frac{\omega x}{2}\right) {\rm Ci}(\omega x)
+\sin\left(\frac{\omega x}{2}\right)\right] \frac{\ell_{\rm P}^4}{4\epsilon^2}\left[\frac{5 \rho^2}{2}+3\sigma\right]
\nonumber
\end{eqnarray}
%\begin{eqnarray}\label{sturmsol}
% \phi(x,w) &=& {\frac {\sin \left( \frac{\omega\,x}{2} 
%\right) }{x}}+
%\frac{1}{3x^3}\, \left(
%-{\frac {5}{16}}\,{\frac {{\rho}^{2}{\ell_{\rm P}}^{4}(1-2\Lambda)}{{
%\epsilon}^{2}}}-\frac{3}{8}\,{\frac {\sigma\,{\ell_{\rm P}}^{4}(1-2\Lambda)}{{
%\epsilon}^{2}}} \right)\times\\
%&&\times \left\{ \cos \left( \frac{\omega\,x}{2} \right) 
%\omega\,x+2\,\sin \left( \frac{\omega\,x}{2} \right) {\it Ci} \left( \omega
%\,x \right) {x}^{2}{\omega}^{2}-2\,\sin \left(\frac{\omega\,x}{2} \right) 
%-2\,\cos \left( \frac{\omega\,x}{2} \right) {\it Si} \left( \omega\,x
% \right) {x}^{2}{\omega}^{2} \right\}  \nonumber\\
%\phi(x,t)&=&\int_0^\infty d\omega \phi(x,\omega) 
%\left( C(\omega){e^{i(1-2\Lambda)\,\omega\,t}}
%+ \bar{C}(\omega)
%{e^{-i(1-2\Lambda)\,\omega\,t}} \right)\nonumber
%\end{eqnarray}
and this solution neglects terms with higher powers than $\ell_{\rm P}^4/(\epsilon x)^2$.
Where 
${\rm Si}(x)\equiv\int_0^x dt\, {\sin(t)}/{t}$, and
${\rm Ci}(x)\equiv \gamma +\ln(x) + \int_0^x dt {(\cos(t)-1)}/{t}$
are the sine integral and cosine 
integral functions respectively and Euler's Gamma is given by 
$\gamma=0.5772156649$. The first term in the bracket in (\ref{sturmsol})
corresponds to the standard spherical mode decomposition in (locally) 
flat space-time.
The next parenthesis includes two terms that are corrections, the first due
to polymerization and the next, involving $\sigma$ is a quantum correction.
These terms would not be present in a treatment of quantum field theory
on a classical space-time. Using the Hamilton equations we can compute
$P^\varphi$,
\begin{equation}
P^\varphi(x,t) = 
\frac{x^2 g(x)^2}{2\sqrt{\omega}(1-2\Lambda)}\frac{\partial\phi(x,t)}{\partial t}
%\frac{x^2 g(x)^2}{2(1-2\Lambda)}\frac{\partial \phi(x,t)}{\partial t}
\end{equation}
and use it to compute the effective Hamiltonian (\ref{effectiveham}),
\begin{equation}\label{effectivehamfock}
  \hat{H}^{\rm eff}_{\rm matt} = (1-2\Lambda)\int_0^{2\pi/\epsilon}d\omega
\omega \hat{\bar{C}}(\omega) \hat{C}(\omega).
\end{equation}
To obtain this expression we note that the solution (\ref{sturmsol})
can be written as $\phi(x,t)=\int_0^\infty d\omega u(x,\omega) h(\omega,t)$ where
$h(\omega,t)$ is the last parenthesis in (\ref{sturmsol}). Notice that
we have introduced a lattice cutoff for the frequency $2\pi/\epsilon$. 
Then one uses the lattice version of the 
closure relation $\int_0^\infty d\omega u(x,\omega)
u(x',\omega)=2 \delta(x-x')/A(x)$ and the orthogonality relation
$\int_0^\infty dx A(x) u(x,\omega)
u(x,\omega')/2=\delta(\omega-\omega')$. 

We have therefore concluded the computation of the state that we will
use as a trial in the variational method. It will be given by a direct
product of the vacuum of the matter part of the Hamiltonian
(\ref{effectivehamfock}) and the Gaussian (\ref{state}) on the
gravitational variables. 
\begin{equation}
  \vert \psi^{\rm trial}_{\vec{\sigma}}\rangle = \vert \psi_{\vec{\sigma}}\rangle \otimes\vert 0\rangle
\end{equation}
The parameters $\vec{\sigma}$ will be varied to minimize the master constraint.
Notice that the state is a direct product
because we are considering the vacuum. If we were to consider
excitations then there might be entanglement between the matter and
gravitational variables \cite{husainterno}.

\subsection{Minimizing the master constraint}

The realization of the master constraint (\ref{masterconstraint}) as a
quantum operator depends on the realization of six key operators. We
proceed to present their expectation values here. We start by the
operators involving the cosine of $\hat{K}_\varphi$,
\begin{eqnarray}\label{68}
\langle \psi^{\rm trial}_{\vec{\sigma}} \vert 
\cos\left(2 \rho \hat{K}_\varphi(i)\right)\vert 
\psi^{\rm trial}_{\vec{\sigma}} \rangle &=& 
\exp\left(-\frac{2 \rho^2}{\sigma(i)}\right),\\
\langle \psi^{\rm trial}_{\vec{\sigma}} \vert 
\cos\left(4 \rho \hat{K}_\varphi(i)\right)\vert 
\psi^{\rm trial}_{\vec{\sigma}} \rangle &=& 
\exp\left(-\frac{8 \rho^2}{\sigma(i)}\right).
\end{eqnarray}

We then consider the powers of the inverse of $\hat{E}^\varphi$. We already
computed the expectation value of the square in (\ref{expectunosobreE}). 
Here we list the other needed powers,
\begin{eqnarray}
&&\langle \psi^{\rm trial}_{\vec{\sigma}} \vert 
  \frac{1}{\left(\hat{E}^\varphi(i)\right)^4}
\vert \psi^{\rm trial}_{\vec{\sigma}} \rangle =
{\frac {(1-2\,\Lambda)^2}{{\epsilon}^{4}{x(i)}^{4}}}+
\frac{5}{4}\,{\frac {{{\ell_{\rm P}}}^
{4} \left( 1-2\,\Lambda \right) ^{3}{\rho}^{2}}{{\epsilon}^{6}{x(i)}^{6}}
}+\frac{5}{2}\,{\frac {\sigma\,{{\ell_{\rm P}}}^{4} \left( 1-2\,\Lambda \right) ^{3}
}{{\epsilon}^{6}{x(i)}^{6}}},\\
&&\langle \psi^{\rm trial}_{\vec{\sigma}} \vert 
\frac{1}{\hat{E}^\varphi(i)}
\cos\left(2\rho \hat{K}_\varphi(i)\right)
\frac{1}{\hat{E}^\varphi(i)}
\vert \psi^{\rm trial}_{\vec{\sigma}} \rangle =
\frac{1-2\Lambda}{\epsilon^2 x(i)^2 
\exp(\frac{2\rho^2}{\sigma})}
\left(1+\frac{5}{2}\frac{\rho^2 l_{p}^{4}}
{\epsilon^2 x(i)^2} + \frac{3}{4}\frac{\sigma l_{p}^{4}}
{\epsilon^2 x(i)^2} \right),\\
&&\langle \psi^{\rm trial}_{\vec{\sigma}} \vert 
\frac{1}{\left(\hat{E}^\varphi(i)\right)^{3/2}}
\sin\left(\rho \hat{K}_\varphi(i)\right)
\frac{1}{\left(\hat{E}^\varphi(i)\right)^{3/2}}
\vert \psi^{\rm trial}_{\vec{\sigma}} \rangle = 0,\\
&&\langle \psi^{\rm trial}_{\vec{\sigma}} \vert 
\frac{1}{\sqrt{\hat{E}^\varphi(i)}}
\sin\left(\rho \hat{K}_\varphi(i)\right)
\frac{1}{\sqrt{\hat{E}^\varphi(i)}}
\vert \psi^{\rm trial}_{\vec{\sigma}} \rangle = 0,\\
&&\langle \psi^{\rm trial}_{\vec{\sigma}} \vert 
\frac{1}{\sqrt{\hat{E}^\varphi(i)}}
\sin\left(3 \rho \hat{K}_\varphi(i)\right)
\frac{1}{\sqrt{\hat{E}^\varphi(i)}}
\vert \psi^{\rm trial}_{\vec{\sigma}} \rangle = 0.
\label{74}
\end{eqnarray}

With these results we can proceed to compute the
expectation value of the master constraint on the gravitational state.
The result will be an operator acting on the matter part.  The
calculation of the expectation values of the coefficients
$\hat{c}_{i}$ and $\hat{c}_{ij}$ (\ref{44})-(\ref{53}) is
straightforward, but lengthy. We will not list the results here.  What
is more challenging is the computation of the expectation value of the
matter part of the expansion of (\ref{masterconstraint}). It helps
that some of the coefficients vanish. The non-vanishing contributions
are,
\begin{eqnarray}\label{masterconstraint2}
\langle \psi_{\vec{\sigma}} \vert 
\hat{\mathbb H}(i) 
\vert \psi_{\vec{\sigma}} \rangle 
&=& 
\ell_{\rm P}\left[
\langle \hat{c}_{11}(i)\rangle \left(\widehat{H_{\rm matt}^{(1)}(i)}  \right)^2
+\langle \hat{c}_{22}(i)\rangle \left(\widehat{H_{\rm matt}^{(2)}(i)}  \right)^2 
+ \langle \hat{c}_{1}(i)\rangle\widehat{H_{\rm matt}^{(1)}(i)}  
+ \langle \hat{c}_{33}(i)\rangle 
\left(\widehat{H_{\rm matt}^{(3)}(i)}\right)^2\right.\nonumber\\
&&\left.+ \langle \hat{c}_{3}(i)\rangle \widehat{H_{\rm matt}^{(1)}(i)} 
+ \langle \hat{c}_{13}(i)\rangle \widehat{H_{\rm matt}^{(1)}(i)} \widehat{H_{\rm matt}^{(3)}(i)}
+ \langle \hat{c}_{00}(i)\rangle\right].
\end{eqnarray}

We now need to compute the expectation value of this operator on the
matter vacuum. To do this we again use the procedure of going to the
continuum limit in the matter terms involving derivatives and momenta
and integrating in the frequencies with an ultraviolet cutoff. Let us
start with $H^{(1)}_{\rm matt}(i)$. The continuum limit expression is
$H^{(1)}_{\rm matt}(x,t)=\ell_{\rm P}^2\left(\left(P^\phi(x,t)\right)^2 +x^4
\left(\phi'(x,t)\right)^2\right)$. We now substitute $P^\phi$ and $\phi$ by 
their mode decomposition and one gets a quadratic expression in the 
$\hat{C}$'s and $u's$. The expectation value only gets contributions
from the $\hat{C}\hat{\bar{C}}$ terms. The result is,
\begin{equation}
  \langle 0 \vert \hat{H}^{(1)}_{\rm matt} \vert 0\rangle 
=l_{p}^{2}\int_{0}^{\frac{2\pi}{\epsilon}}d\omega 
\frac{1}{8\omega(1-2\Lambda)}
[A(x)^2u^2(x,\omega)\omega^2(1-2\Lambda)^2+4x^4(\partial_{x}u(x,\omega))^2],
\end{equation}
and substituting $u(\omega,x)$ and $A(x)$ we obtain,
\begin{eqnarray}
  \langle 0 \vert \hat{H}^{(1)}_{\rm matt}(x) \vert 0\rangle &=&
l_{p}^{2}(1-2\Lambda)A(x)^2\left(\frac{\pi^2}
{8x^2\epsilon^2}+\frac{1}{8x^4}-
\frac{\cos^2(\frac{\pi x}{\epsilon})}{8x^4}- 
\frac{\pi\sin(\frac{\pi x}{\epsilon})\cos(\frac{\pi x}{\epsilon})}
{4x^3\epsilon} \right)\nonumber\\
&&+\frac{l_{p}^{2}}{(1-2\Lambda)}
\left(\frac{\pi^2 x^2}{8 \epsilon^2}+\frac{\ln(2)}{4}
+\frac{x\pi\cos(\frac{\pi x}{\epsilon}) 
\sin(\frac{\pi x}{\epsilon})}{4\epsilon} - 
\frac{5}{8}\sin^2(\frac{\pi x}{\epsilon}) +
\frac{1}{4}{\rm Cin}(\frac{\pi x}{\epsilon}) \right),
\end{eqnarray}
where ${\rm Cin}(x)=\gamma+\ln x -{\rm Ci}(x)$. One can get a more manageable
expression, which we will use in the rest of the paper by 
ignoring corrections of $\ell_{\rm P}^4$ and neglecting 
the highly oscillating terms that involve $\sin(\pi x/\epsilon)$ or
cosines and the integral cosines. The result is,
\begin{equation}
  \langle 0 \vert \hat{H}^{(1)}_{\rm matt}(x) \vert 0\rangle =
\frac{l_{p}^{2}}{4(1-2\Lambda)}
\left(-2+\frac{\pi^2 x^2}{\epsilon^2}
+\ln(2)+\gamma+\ln(\frac{\pi x}{\epsilon})\right),\label{cr:78}
\end{equation}
and the dominant term is $\pi^2 x^2/\epsilon^2$. Reverting to the 
discrete theory, it reads,
\begin{equation}
  \langle 0 \vert \hat{H}^{(1)}_{\rm matt}(i) \vert 0\rangle =
\frac{l_{p}^{2}\epsilon^3}{4(1-2\Lambda)}
\left(-2+\frac{\pi^2 x(i)^2}{\epsilon^2}
+\ln(2)+\gamma+\ln(\frac{\pi x}{\epsilon})\right).
\end{equation}
The procedure to compute the expectation value of the other terms in
(\ref{masterconstraint2}) is exactly the same, but the size of the expressions
involved is quite large. We will not display them here for reasons of space.

The result for the expectation value of the 
integrand of the master constraint is,
\begin{eqnarray}
\langle \hat{\mathbb H}(x) \rangle &=&
  {\frac {\sigma_0\,{\ell_{\rm P}}^{3}}{\epsilon\,{x}^{2}}}+ \left( 
8\,{\frac {{\pi }^{2}}{{\epsilon}^{3}{x}^{2}}}
+\frac{32}{\epsilon x^4}\,\ln  \left( {\frac {L}{\epsilon}} \right) 
-\frac{\left(\gamma-2+\ln\left(\frac{2\pi x}{\epsilon}\right)\right)\pi}
{\epsilon\,{x}^{4}(1-2\Lambda)}
+{\frac {1}{96}}\,{\frac {{\pi }^{3}}{{\epsilon}^{5}{x}^{2}\sigma_0\,
{(1-2\Lambda)}^{2}}}\right.\\
&&
-\frac{1}{48}\,{\frac {\Lambda
\,{\pi }^{3}}{{\epsilon}^{5}{x}^{2}\sigma_0\,{(1-2\Lambda)}^{2}}}
-{\frac {43}{128}}\,{\frac {\Lambda\,\pi }{{\epsilon}^{3}{x}^{4}
\sigma_0\,{(1-2\Lambda)}^{2}}}\nonumber\\
&&
+{\frac {\epsilon\,(\gamma-2+\ln  \left(\frac{2\pi x}{\epsilon} \right)) 
\pi }{{x}^{4}
(1-2\Lambda)\,{L}^{2}}}
+8\,{\frac {\epsilon\,{\pi }^{2}}{{x}^{2}{L}^{4}}}
-\frac{2\,\pi}{\epsilon x^4 (1-2\Lambda)} 
\,\ln  \left( {\frac {L}{\epsilon}} \right)
-16\,{\frac {{\pi }^{2}}{\epsilon\,{x}^{2}{L}^{2}
}}
+\frac{1}{32}\,{\frac {\pi }{\epsilon\,{x}^{4}{(1-2\Lambda)}^{2}}}
\nonumber
\\
&&+\frac{1}{48}\,{\frac {{\pi }^{3}}{{\epsilon}^{3}{x}^{2}{(1-2\Lambda)}^{2}}
}
+\frac{32\,\epsilon}{x^6\pi^2}
\, \left( \ln  \left( {\frac {L}{\epsilon}} \right) 
 \right) ^{2}
-{\frac {{\pi }^{3}}{{\epsilon}^{3}{x}^{2}(1-2\Lambda)}
}
+4\,{\frac {\pi }{\sigma_0\,{\epsilon}^{3}{x}^{2}}}\nonumber\\
&&
-\frac{32}{x^4 L^2}\,\epsilon
\,\ln  \left( {\frac {L}{\epsilon}} \right) -3\,{
\frac {\sigma_0\,\pi }{{x}^{3}{L}^{2}}}+{\frac {43}{256}}\,{\frac 
{\pi }{{\epsilon}^{3}{x}^{4}\sigma_0\,{(1-2\Lambda)}^{2}}}
+\frac{8}{\sigma_0\epsilon x^4 \pi}\,
\ln  \left( {\frac {L}{\epsilon}} \right) 
\nonumber\\
&&-\frac{1}{4(1-2\Lambda)}\frac{\left(4x^2\epsilon\sigma_0
+4\,x{\sigma_0}^{3}{\epsilon}^{2}+4\,\sigma_0\,{\epsilon}^{2}
x+7\,{\sigma_0}^{3}{\epsilon}^{3} \right)}{(x+\epsilon)\sigma_0^2 x^5 \epsilon^4}
\times \left( 
\frac{1}{4}\pi^2 x^2
+\frac{1}{4}\epsilon^2\left(\gamma-2+\ln\left(\frac{2\pi x}{\epsilon}\right)
\right) \right)\nonumber\\
&&-3\,{\frac {\sigma_0\,\pi }{ \left( x+\epsilon
 \right) {x}^{2}{\epsilon}^{2}}}+3\,{\frac {\sigma_0\,\pi }{
 \left( x+\epsilon \right) {x}^{2}{L}^{2}}}
-\frac{6\,\sigma_0}{(x+\epsilon)x^4 \pi}
\,\ln\left( {\frac {L}{\epsilon}} \right)  
-4\,{\frac {\pi }{{x}^{2}\epsilon\,\sigma_0\,{L
}^{2}}}
+{\frac {{\pi }^{3}}{\epsilon\,{x}^{2}(1-2\Lambda)\,{L}^{2}}}\nonumber\\
&&\left.
-2\,\frac{\epsilon}{x^6(1-2\Lambda)\pi}
\left(\gamma-2+
\ln\left( {\frac {2\pi x}{\epsilon}}\right)\right) 
\ln\left( {\frac {L}{\epsilon}}\right)
+3
\,{\frac {\sigma_0\,\pi }{{x}^{3}{\epsilon}^{2}}}
+6\,\frac{\sigma_0}{x^5\pi}
\,\ln  \left( {\frac {L}{\epsilon}} \right) 
-1/16\,
{\frac {\epsilon}{{x}^{6}{(1-2\Lambda)}^{2}\pi }}
 \right) {\ell_{\rm P}}^{5}\nonumber.
\end{eqnarray}
We have assumed $\sigma=\sigma_0 \epsilon^2/\ell_{\rm P}^2$ with $\sigma_0$
of order unity and we have neglected terms $O(\ell_{\rm P}^7)$. We have
assumed $\sigma$ to be independent of $x$ in order to simplify the
above expression, which otherwise becomes too large. Experiments we
have carried out suggest that allowing variations in $x$ leads to
the same minimum value of $\sigma$ approximately independent of $x$.

We would like to study the minimum of the master constraint as a function
of $\sigma_0$ for different choices of $\epsilon/\ell_{\rm P}$. Notice that
we have assumed $\sigma_0$ to be of order one. One can change that by
varying the ansatz for $\sigma$ including other powers $\epsilon/\ell_{\rm P}$
different than 2. We have carried out such experiments. The results
can be summarized as follows.
\begin{figure}
\includegraphics[height=5cm]{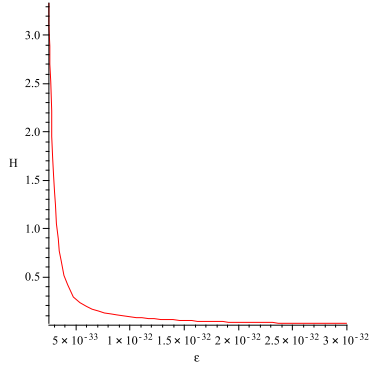}
\caption{The expectation value of the
master constraint as a function of the lattice spacing.
We see that the value of the master constraint is small unless
one chooses lattice separations of order Planck length. The figure does not
show it, but for separations of the order of $10^{-23}$cm the master constraint is very small, of the order of $10^{-20}$ (we are using units in which 
$\hbar$ is one and therefore the master constraint is dimensionless).}
\label{figure1}
\end{figure}
In figure \ref{figure1} we show the value of the master constraint as
a function of $\epsilon$ (in centimeters) and for $\sigma_0=10$ and
$\sigma =\sigma_0 \epsilon^3/\ell_{\rm P}^3$. 
Varying $\sigma_0$ while keeping it of order one changes little the
shape of the curve. We see that in the approximation studied the
theory does not appear to have a continuum limit, but we see that the
master constraint quickly drops to zero for lattice spacings larger
than Planck scale. Although the figure suggests that the master
constraint drops even further for larger lattice spacings, the
approximation in which we have handled expressions (in which we have
neglected higher powers of $\epsilon/\ell_{\rm P}$) is inadequate for large
values of $\epsilon$ and the master constraint very likely will
increase its value for large values of $\epsilon$. So there exists a
genuine preferred value of $\epsilon$ that minimizes the master
constraint.  Even so, the approximation should be reliable up to
values of $\epsilon\sim 10^{-23}$cm  and for such values the master
constraint is of the order of $10^{-20}$, so one sees that this is a
regime where one approximates the continuum theory very well.

We have explored other ranges of $\sigma$'s (with different powers of
$\epsilon/\ell_{\rm P}$). The observation is the following. For lower
powers than three we get a curve that looks similar to the one shown
in the figure, but that grows faster as one approaches smaller lattice
spacings and therefore the minimum occurs farther away from the Planck
scale. For powers higher than $10/3$ one violates the approximation
that $\ell_{\rm P}/\epsilon$ is small and the expressions we derived
are not valid. From these considerations and an analysis of the powers
involved, we conclude that the minimum for the master constraint is
achieved for a power of $\epsilon/\ell_{\rm P}$ in $\sigma$ close to
two and $\epsilon \sim 10^{13}\ell_{\rm P}$.

An interesting speculation is that if the minimum of the master
constraint happens in the range mentioned, the cosmological constant,
which goes as $\Lambda\sim \ell_{\rm P}^2/\epsilon^2$ would not be of Planck
scale but several orders of magnitude smaller.

Another observation of interest is to note what would have happened if
instead of choosing the state peaked around the flat metric (with a
topological defect) one would have chosen the ``loop quantum gravity
vacuum'', i.e. a state with zero loops which corresponds to a
degenerate metric $\vert \mu(i)=0\rangle$. Such a state annihilates
the matter Hamiltonian in the loop representation and has zero volume.
It would be disturbing if this state yielded a lower value for the
master constraint than the state we constructed, since it would imply
that degenerate geometries dominate. This is not the case, as can be
easily seen. For such a state all expectation values
(\ref{68})-(\ref{74}) vanish. One can check that the 
expectation value of the master constraint is,
\begin{equation}
\langle \hat{\mathbb H} \rangle =  \frac{1}{8} \frac{L \ell_{\rm P}}
{\epsilon^2 \rho}.
\end{equation}
That is, the result is very large. For $\epsilon\sim \ell_{\rm P}$ it goes
as $L/\ell_{\rm P}$, the size of the universe in Planck lengths. Therefore
these degenerate states are heavily suppressed.

\section{Discussion}

We have studied spherically symmetric gravity coupled to a spherically
symmetric scalar field using loop quantum gravity techniques. The
problem has a non-Lie algebra of constraints and we used the ``uniform
discretization'' technique to treat the dynamics. We used a
variational technique to minimize the discrete master constraint.
With the trial states proposed, we were not able to reach a zero eigenvalue
for the master constraint, that is, the theory does not seem to
have a quantum continuum limit. The lowest eigenstate of the master
constraint has the form of a direct product of a Fock vacuum for the
scalar field and Gaussian states centered around flat space-time for
the gravitational variables. Although the theory does not have a
continuum limit, it approximates general relativity well for small
values of the lattice separation, which in turn regularizes the
cosmological constant. The lattice treatment we have performed diverges when
one takes the continuum limit. The reader may wonder why loop quantum gravity
has failed to act as the ``natural regulator of matter quantum field
theories'' as claimed, for instance in \cite{qsd5}. The problem arises
with the gauge fixing of the diffeomorphism constraint that we
performed at the classical level. This leads us to variables that have
the structure of a Bohr compactification in the ``transverse''
$\varphi$ direction, but the variable in the radial direction is a
c-number and therefore is not dynamical and has continuous character.
There is no chance therefore that loop quantum gravity based on this
gauge fixing could regulate the short distance behavior, which is
responsible for the emergence of the cosmological constant. To tackle
this issue one would have to allow both the diffeomorphism and
Hamiltonian constraint to remain in the theory. The calculational
complexity would increase importantly, since one will have to regulate
the master constraint in such a way that the resulting states have
remnants of diffeomorphism invariance in the discrete theory. This has
been successfully accomplished with uniform discretizations in the
Husain--Kucha\v{r} model \cite{difeos}, but the complexity there was
considerably reduced by the lack of a Hamiltonian constraint. It is
worthwhile noticing that even if one allowed loop quantum gravity to
regulate matter in the proposed way, the resulting cosmological
constant is likely to be finite but still very large with respect to
the current observed value.

The present paper is a first exploration of a difficult problem,
carried out with several assumptions and limitations that we have
outlined in the text.  Future work will include relaxing the
assumption that one has a Fock vacuum for the scalar field and
treating both the gravitational and scalar variables on the same
footing with the variational technique for the master constraint. In
this context it will be interesting to study the excited states of
matter and study the modifications in dispersion relations for the
matter fields due to the quantum geometry.  This will definitely
require considering trial states with correlations in the variational
method, something we have not done here. One should also relax the
gauge fixing of diffeomorphisms to see if the cosmological constant
problem becomes better under control. Other future directions
would be to consider solutions centered around non-flat geometries,
for instance, including a black hole with the aim of studying if the
scalar field states involve Hawking radiation.

\section*{Acknowledgments}

 This work was supported in part by
grant NSF-PHY-0650715, funds of the Hearne Institute for Theoretical
Physics, FQXi, CCT-LSU, Pedeciba and ANII PDT63/076.

\end{document}